\documentstyle[prl,aps,epsfig,multicol]{revtex}
\begin{document}
\draft
\title{Driven Tunneling Dynamics: Bloch-Redfield Theory versus Path
Integral Approach}
\author{Ludwig Hartmann$^1$, Igor Goychuk$^1$, Milena Grifoni$^{2,3}$ and
Peter H\"anggi$^1$}
\address{$^1$Institut f\"ur Physik,
         Universit\"at  Augsburg, Universit\"atsstr.\ 1, D-86135 Augsburg,
         Germany\\
         $^2$Institut f\"ur Theoretische Festk\"orperphysik, Universit\"at
         Karlsruhe, D-76128 Karlsruhe, Germany\\
         $^3$Dipartimento di Fisica, INFM, Via Dodecaneso 33, I-16146,
Genova, Italy\\  }
\date{Date: \today}
\maketitle
\widetext
\begin{abstract}
In the regime of weak bath coupling and low temperature we
demonstrate numerically for the spin-boson dynamics the
equivalence between two widely used but seemingly different
roads of approximation, namely the path integral approach
and the Bloch-Redfield theory. The excellent agreement between
these two methods is corroborated by a novel efficient {\it analytical}
high-frequency approach: it well approximates the decay of quantum coherence
 via a series of damped coherent oscillations. Moreover,
a suitably tuned control field can  selectively enhance or
suppress quantum coherence.
\end{abstract}
\vspace{5mm}
\raggedcolumns
\begin{multicols}{2}
\narrowtext
\noindent

The dynamics of driven  quantum systems which interact with a large number
of environmental
degrees of freedom \cite{Leggett87,Weiss93,physrep}
plays an increasingly
prominent role: its vast applicability ranges from tunneling phenomena in
solid state physics,
the study of electron and proton transfer in condensed phases,  to the
gate operation
in quantum computing devices \cite{quantComp}, to name but a few. In
particular, the use of
properly tailored
external driving forces enables one to selectively manipulate a quantum
transport process.
The various communities  typically rely on different methods of description.
The two most popular approaches for a portrayal of the time evolution of the
corresponding reduced density matrix (RDM) are either based on the
system-bath coupling expansion
obtained by use of a projector operator method -- being
commonly known under the label of the {\em Bloch-Redfield formalism} -- or on the
expansion in the
coupling matrix element $\Delta$ (such as a tunnel splitting) by use of
(real-time) {\em path integral methods}.
Nevertheless, there exists practically
little crosstalk between the
practitioners of the two approaches, and  even more, not much  of detailed
comparison between the two seemingly
different roads of approximation needed for practical calculations.

For the archetype quantum system of a driven spin-boson dynamics, namely
the driven
dissipative two-state system (TSS) dynamics \cite{physrep}, the application of
the  so termed
noninteracting blip approximation (NIBA), i.e. the leading order result in
the tunnel coupling $\Delta^2$,  produced many impressive successes
in entangling the complexity
of  driven open quantum systems. This scheme works best in the regime of
strong friction and/or high thermal
temperatures. Much less is presently known, however, about the corresponding
complexity of the driven dynamics
in the deep quantum regime at low temperatures and weak system-bath
coupling,  where the NIBA
is failing and higher order terms in the series in $\Delta$ must be
accounted for \cite{Gri96,Sigmax}. In practice, this latter regime
is of relevance for many situations such as, e.g., for the challenge of
"battling decoherence" in quantum
computing schemes \cite{quantComp}.

Our main objective with this work is to enlight the advantages and
disadvantages of the two approaches.
In doing so we present three major findings: (i) We numerically demonstrate the
 equivalence for the
 driven tunneling dynamics between the   path integral
method beyond NIBA
and the coupled set of nonstationary, Markovian Bloch-Redfield equations.
(ii) Starting from
the generalized master equation
(GME) for the RDM, obtained within the path integral approach, we arrive at
a novel {\it analytic} high
driving frequency approximation that compares well with comprehensive
numerical findings. (iii) With this
analytical result one can efficiently determine the optimal control of
quantum coherence.

Our starting point is the  {\it driven} spin-boson
Hamiltonian \cite{physrep}
where the TSS is bilinearly coupled to an ensemble
of harmonic oscillators, i.e.,
\begin{eqnarray}\nonumber
\hat H(t)&=& -\hbar[\Delta \hat \sigma_x  +\epsilon(t)\hat \sigma_z]/2 \\
\vspace{-1mm}
&+& \sum_{i}\hbar\omega_i ( \hat b_i^\dagger \hat b_i
+ 1/2)+\hat\sigma_z \sum_{i}  c_i (\hat b_i+\hat b_i^\dagger)/2 \;,
\label{Hamilton}
\end{eqnarray}
with $\hat\sigma_i$ being Pauli spin matrices.
Here $\Delta$ describes the coupling between the two states, and $\epsilon(t)$
is the external, time-dependent control field.
The basis states are chosen such that
$|R\rangle$ (right) and $|L\rangle$ (left)
are the localized eigenstates of the "position" operator $\hat \sigma_z$.
All effects of the Gaussian  bath on the TSS are captured by the
force autocorrelation function
\cite{Leggett87,Weiss93,physrep}
${\cal M}(t)=\frac{1}{\pi}\int_0^{\infty} d\omega
J(\omega) \; \frac{\cosh(\hbar\omega/2k_BT - i\omega t)}
{\sinh(\hbar\omega/2k_BT)},$
where the spectral density of the environment, $J(\omega)=
\pi\hbar^{-2}\sum_{i} c_i^2 \delta(\omega-\omega_i)=2\pi\alpha\omega e
^{-\omega/\omega_c}$,
is assumed to be of Ohmic form with exponential cutoff and dimensionless
coupling strength $\alpha$.
The dynamical quantities of interest are the expectation values
$\sigma_i(t) := {\rm Tr}\{\hat \rho(t)\hat \sigma_i\}$
which, together with the unit matrix $\hat I$, comprise
the complete reduced density matrix
$\hat\rho(t) =  \hat I/2 + \sum_{i=x,y,z}\sigma_i(t)\hat \sigma_i/2$.
In the following we assume
that at time $t=0$ the particle is held at the right site
$\sigma_z=+1$, with the bath being in thermal equilibrium.

{\em Path-Integral Approach.} -- For a harmonic bath
the exact formal solution for the evolution of the $\sigma_i(t)$ can be
expressed in  terms of  real-time double path integrals
\cite{Leggett87,Weiss93,physrep}.
This procedure yields
the formally exact set of equations \cite{physrep,Gri96,Sigmax}
\vspace{-1mm}
\begin{eqnarray}\label{GME}
\dot\sigma_z(t)&=&\int_{0}^{t} dt'
\left[K_z^{(-)}(t,t')-K_z^{(+)}(t,t')\;\sigma_z(t')\right]\;, \\
\sigma_x(t)&=&\int_0^tdt'\left[K_x^{(+)}(t,t')+K_x^{(-)}(t,t')\sigma_z(t')\right
]\;, \nonumber
\vspace{-1mm}
\end{eqnarray}
and $\sigma_y(t)=-\dot\sigma_z(t)/\Delta$.
Here, the  kernels $K^{(\pm)}_i$, $i=x,z$ are
found in the form of a series expansion in $\Delta$.
Because the exact series expression cannot be evaluated to all orders,
approximation schemes necessarily must be invoked.
A familiar scheme is the
noninteracting-blip approximation (NIBA)
\cite{Leggett87,Weiss93,physrep}, which
corresponds to a truncation of the series expansion
to lowest order in $\Delta$.
The NIBA is approximatively valid only for the dynamics of
$\sigma_z(t)$ if on average $\langle\epsilon (t) \rangle = 0$.
However, in the presence of a static asymmetry component,
it breaks down for weak
damping {\em and} low temperatures \cite{Weiss93,physrep}.
A systematic weak damping approximation
for the kernels $K_i^{(\pm)}$ in  (\ref{GME}),
which circumvents the weaknesses of the NIBA has been discussed in
\cite{Gri96,Sigmax}.
By  keeping track of the bath-induced correlations
to {\em linear} order in $\alpha$,
the whole series expansion in $\Delta$ can be summed analytically.
The kernels $K_z^{(\pm)}$ read
\vspace{-1mm}
\begin{eqnarray}
K_z^{(+)}(t,t')&=&\Delta^2 \cos[\zeta(t,t')] [1-Q'(t-t')] \nonumber \\
+ \int_{t'}^{t} &dt_2& \int_{t'}^{t_2} dt_1 \Delta^4 \sin[\zeta(t,t_2)]
P_0(t_2,t_1)\sin[\zeta(t_1,t')] \nonumber \\
\times [Q'(t&-&t')+Q'(t_2-t_1)-Q'(t_2-t')-Q'(t-t_1)]\;,\nonumber\\
\nonumber \\
\vspace{-1mm}
K_z^{(-)}(t,t')&=&\Delta^2 \sin[\zeta(t,t')] Q''(t-t') \nonumber \\
-\int_{t'}^{t} &dt_2& \int_{t'}^{t_2} dt_1 \;\; \Delta^4 \sin[\zeta(t,t_2)]
P_0(t_2,t_1)\cos[\zeta(t_1,t')] \nonumber \\
\times [Q''(t&-&t')-Q''(t_2-t')]\;. \label{kernel-a-weak}
\vspace{-1mm}\end{eqnarray}
Here, the first term in $K_z^{(\pm)}$  represents the 
weak-coupling form of NIBA.
In the remaining contribution the term $P_0(t_2,t_1)$
accounts for all tunneling events during the
time interval $[t_1,t_2]$ that are not influenced by damping.
Hence $P_0(t_2,t_1)$ solves the generalized master equation (GME) for
$\sigma_z(t)$ (\ref{GME}) with the zero-damping kernels
  $K^{(+)}(t,t')=\Delta^2\cos[\zeta(t,t')]$ and
  $K^{(-)}(t,t')=0$,
where $\zeta(t,t')=\int_{t'}^t dt'' \epsilon(t'')$
captures the effects of the external force.
The bath-influence is encapsulated
in the functions
$Q'(t)$ and $Q''(t)$ being the
real and imaginary part, respectively, of the twice integrated
bath correlation function ${\cal M}(t)$.

{\em Bloch - Redfield Formalism.} -- In the Nakajima-Zwanzig
theory \cite{Nakajima} it is well known how
to construct an exact generalized master equation 
for the reduced density matrix
with the help of projection operators.
For intermediate to high temperatures and/or
strong damping, but for arbitrary driving, a  master equation
for $\hat\rho(t)$
can be obtained  within  the small
polaron theory, yielding
equations which are equivalent to the NIBA  \cite{Goy,rem2}.
For weak coupling to the bath the projection
operator technique yields the GME in Born approximation that
can be further simplified to the Markovian kinetic equations
without loss of accuracy to the leading order in dissipative coupling.
For strong harmonic driving this objective was first achieved
in 1964 by Argyres and Kelley \cite{Argyres}.
Following the reasoning in \cite{Argyres} the kinetic equations for the
RDM of a
stochastically driven TSS were found in [12a] and in a different way in
[12b].
Generalizing \cite{Argyres,Igor95} to
the case of a spin-boson problem with an {\it arbitrary} control field
we find the
coupled equations
\vspace{-1mm}
\begin{eqnarray}
\dot \sigma_x (t)&=& \epsilon(t)\sigma_y-
\Gamma_{xx}(t)\sigma_x-\Gamma_{xz}(t)\sigma_z-A_x(t)
\label{markov}\;,\\
\nonumber\dot\sigma_y(t)&=&-\epsilon(t)\sigma_x+\Delta\sigma_z
- \Gamma_{yy}(t)\sigma_y- \Gamma_{yz}(t)\sigma_z-A_y(t)\;,
\vspace{-1mm}
\end{eqnarray}
with $\Gamma_{yy}(t)=\Gamma_{xx}(t)$
and $\dot \sigma_z= -\Delta\sigma_y$.
Here the time-dependent rates $\Gamma_{ij}(t)=\int_{0}^{t}
dt'{\cal M}'(t-t')b_{ij}(t,t')$,
together with the inhomogeneities
$A_x(t)={\rm Im}F(t)$, $A_y(t)={\rm Re}F(t)$,
with $F(t)=2\int_{0}^{t}dt'{\cal M}''(t-t')
U_{RR}(t,t')U_{RL}(t,t')$
determine the dissipative action of  the thermal bath on the TSS.
The functions  ${\cal M}'$ and ${\cal M}''$
are the real part and imaginary part, respectively,
of the  correlation function ${\cal M}$.
The quantities
$ U_{RR}(t,t')=\langle R| U(t,t') |R\rangle $ and
$U_{RL}(t,t')=\langle R| U(t,t') |L\rangle $ are matrix elements of the
time evolution
operator $U(t,t')$ of the {\em non-dissipative} driven TSS.
The functions $b_{ij}$ read
$b_{xx}=|U_{RR}|^2-|U_{RL}|^2$,
$b_{xz}=2\;{\rm Re}U_{RR}U_{RL}$, and $b_{yz}=-2\;{\rm Im}U_{RR}U_{RL}$.
This main result in
Eq. (\ref{markov}) yields for the first time a consistent
Bloch-Redfield-type description of the externally driven
spin-boson dynamics. Equations of the form (\ref{markov})
were  derived by Bloch and Redfield in 1957 \cite{Bloch57}
to describe spin relaxation in nuclear magnetic resonance,
and in \cite{Aslangul86} for the dynamics of the undriven
spin-boson problem. Our set of
Eqs. (\ref{markov}) generalizes
\cite{Aslangul86} to general driving
forces. Note that these derived equations are valid in the parameter
region $\alpha\ln(\omega_c/\Delta)\ll 1,$ where
the frequency corrections to the dynamics incurred due to the
dissipation are small, and the
perturbative treatment is fair. One can show that
for the undriven case, $\epsilon(t)=\epsilon_0,$  the
analytic solution of Eq. (\ref{markov}) in first order in $\alpha$
reproduces the
analytical path integral weak-damping results, cf. \cite{Weiss93,Sigmax}
and (\ref{P-weak-diss})
below with zero ac-field.

{\em Analytic high frequency solution.} --  Up to here no
assumptions on the deterministic control field have been made.
Next, we focus our attention on a monochromatic field of the form
$\epsilon(t)=\epsilon_0+s\cos\Omega t$. Moreover, we restrict our
investigations on the $\sigma_z (t)$-dynamics, as this quantity is
of prime interest for describing tunneling properties.
\begin{figure}[t]
\epsfig{figure=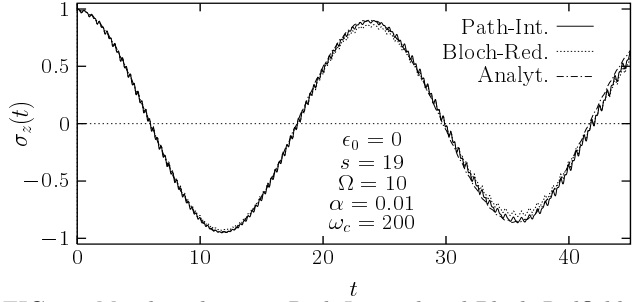,width=8.5cm,height=4cm,angle=0}
\caption{{\em Matching between Path Integral and Bloch-Redfield.} The
comparison of
the dynamical Eqs. (\ref{GME}), (\ref{markov})
and (\ref{P-weak-diss}) for unbiased TSS-dynamics depicts excellent agreement.
For this resonant $(\epsilon_0=n^*\Omega,n^*=0)$
condition the dynamics is well described by (\ref{P-weak-diss}) with the
single-mode frequency $\tilde\theta_0=\tilde\Delta_0$.
Here and in the following figures frequencies are expressed
in units of $\Delta$, times in units of $\Delta^{-1}$. The
temperature is zero throughout.}
\end{figure}
\vspace{-0.2cm}
Because the path integral approach yields a closed integro-differential
equation for $\sigma_z (t)$,
we start from the generalized master equation (\ref{GME}).
In the high frequency regime  ($\Omega\gg \{\Delta,\epsilon_0,\Gamma_{R}\}$,
with $\Gamma_R$ defined in (\ref{gamR}) below)
a good approximation to the
dynamics of $\sigma_z (t)$
amounts to perform the substitution
${\cal R}[\zeta(t,t')] \longrightarrow
 \langle {\cal R}[\zeta(t,t')] \rangle = J_0(\chi)
{\cal R}[\epsilon_0(t-t')]$
into the kernels $K_z^{(\pm)}(t,t')$, where
${\cal R}=\cos$ or $\sin$ and
$\chi=\frac{2s}{\Omega}\sin\frac{\Omega(t-t')}{2}$.
Here $\langle \;\rangle$ denotes time-averaging,
and $J_0$ is the zero order Bessel function.
The resulting generalized master equation
is in the form of a
time-convolution.
A solution is conveniently obtained by use of Laplace transformation
techniques,
upon generalizing a line of reasoning
proposed in Ref. \cite{Gri96}.
By  expanding the  Bessel function $J_0(\chi)$ in Fourier
series, and upon
introducing  the field-dressed tunneling
splittings $\Delta_{n}=|J_{n}(s/\Omega)|\Delta$
and the photon-induced asymmetries $\epsilon_n^{}=\epsilon_0 - n\Omega$,
we end up with
\begin{eqnarray}
\vspace{-3mm}
\sigma_z(t)=P_{\infty}+({\cal P}_0-P_{\infty})e^{-\Gamma_R t}
+ \!\!\!\! \sum_{n=-\infty}^{\infty} \!\!\!\!
{\cal C}_n\cos(\tilde\theta_n t)e^{-\Gamma_n t}\;.
\label{P-weak-diss}
\vspace{-2mm}
\end{eqnarray}
Conservation of probability yields ${\cal P}_0+\sum_n{\cal C}_n=1$, with
${\cal P}_0=\!\!\prod\limits_{n}^{}(\epsilon_n/\theta_n)^2$, and
${\cal C}_n= \!\prod\limits_{m}^{}(\theta_n^2-\epsilon_m^2)/[
\theta_n^2\prod\limits_{m\neq n}{}\!(\theta_n^2-\theta_m^2)]\;.$
The damping rates and the averaged nonequilibrium value $P_{\infty}$ read
\cite{footnote}
\begin{eqnarray}
\vspace{-2mm}
\label{gamR}
\Gamma_R&=&2\sum_{n}^{}\Gamma_n\;,\qquad
\Gamma_n=\frac{1}{4}{\cal C}_n\;f_n^2\;S(\theta_n)\;,\\
\vspace{-2mm}
P_{\infty}&=&\frac{1}{2\Gamma_R}\;\sum_{n}^{}
\sqrt{{\cal P}_0}\;f_n\;{\cal C}_n
J(\theta_n)\sum_{m}^{}\frac{\Delta_m^2}{\theta_n^2-\epsilon_m^2}\;.
\vspace{-2mm}
\end{eqnarray}
Here
$f_n =\sqrt{{\cal P}_0}\;\theta_n\sum_{m}^{}
\Delta_m^2/[\epsilon_m(\theta_n^2-\epsilon_m^2)]$, and
$S(\theta)=J(\theta)\coth(\hbar\theta/2k_BT)$.
The infinite set of
frequencies $\theta_n$ is determined
by the pole equation for the undamped TSS
\begin{equation} \label{pole}
\!\!\!\!\!\!\prod_{n}^{}(\epsilon_n^2-\theta^2)+\sum_{n}^{}
\Delta_n^2\!\!\! \prod_{m;m\neq n}^{}\!\!\!
(\epsilon_m^2-\theta^2)=0\;.
\end{equation}
\begin{figure}[t]
\epsfig{figure=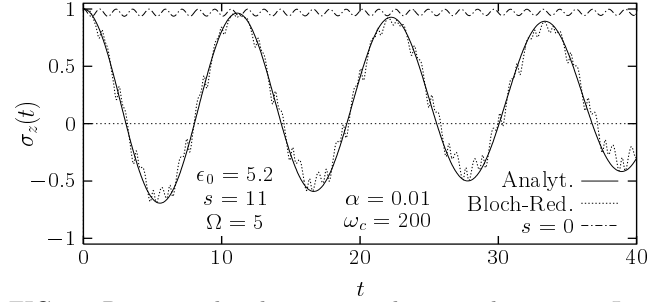,width=8.5cm,height=4cm,angle=0}
\caption{{\em Driving induced quantum coherence phenomena}.
In the presence of a quasi-resonant
high frequency field away from the zeros
of $J_{n^*}(s/\Omega)$,
the population difference $\sigma_z(t)$
exhibits a coherent oscillatory decay which is dominated
by a {\em single } mode oscillation frequency $\tilde\theta_{n^*}$.
A comparison between the predictions of the
analytical solution (\ref{P-weak-diss}) with just the single mode
frequency $\tilde\theta_1$, for a near-resonant field
(i.e., $n^*=1$ with $\epsilon_1=|\epsilon_0-\Omega|=0.2\Delta$)
with the Bloch-Redfield result in (\ref{markov}) is depicted.
Note that in the  undriven situation ($s=0$)
the TSS  dynamics is almost completely {\em localized}.}
\end{figure}
\vspace{-0.2cm}
Finally,
to approximately take into account bath-induced frequency-shifts
the tunneling frequencies $\tilde\theta_n$ are evaluated from
(\ref{pole}) upon substituting
$\Delta_n\to \Delta_n(1-\alpha\ln(\omega_c/\Delta)):=
\tilde\Delta_n$.
Thus, in this
high frequency regime the system generally still exhibits
damped coherent oscillations, as in the undriven case,
although, an {\em infinite} set of
oscillation frequencies $\tilde \theta_n$ with
corresponding damping rates $\Gamma_n$ enters this
driven dynamics. Superimposed to
these  coherent oscillations
there occurs
an incoherent decay with rate $\Gamma_R$ towards
$P_{\infty}$.

In figures 1-3 we depict  comparisons amongst
the numerical predictions
of the Born-Markov equations (\ref{markov}),
the path-integral GME (\ref{GME}), and the analytical
solution (\ref{P-weak-diss}) for small Ohmic friction  and
zero temperature.
For the driven dynamics the agreement is remarkable.
It increases further with increasing temperature (not shown).
To achieve a convergence of (\ref{P-weak-diss}) a
truncation of the pole equation (\ref{pole}) to five (or less, cf. Figs.1,2)
modes, characterized by $(\Delta_n,\epsilon_n)$,
turned out to be sufficient.
In Fig. 1 the influence of an {\em unbiased}
($\epsilon_0=0$) control field
is investigated. This corresponds to a resonant $(\epsilon_0=n^*\Omega)$
field with $n^*=0$.
In Figs. 2 and 3 the case of a
finite bias $\epsilon_0 \neq 0$
is depicted.
Fig. 2  depicts the near-resonant situation
$|\epsilon_0-n^*\Omega|=|\epsilon_{n^*}| \ll\{\tilde\Delta_{n^*},
|\epsilon_0|\}$
away from the zeros of $J_{n^*}(s/\Omega)$:
the coherent dynamics is now already well
captured by the {\em single resonant mode} frequency
$\tilde\theta_{n^*}=\sqrt{\tilde\Delta_{n^*}^2+\epsilon_{n^*}^2}$.
This finding generalizes the small-dc-bias analysis in \cite{Neu}.
In addition, we deduce from the parameters chosen in Fig. 2
that our approach can even work for intermediate driving frequencies
($\Omega\approx\epsilon_0$).
Due to the fact that $\tilde\Delta_{n^*}\le \Delta$, and
$\epsilon_{n^*} <\epsilon_0$,  the field-induced
oscillation frequency $\tilde\theta_{n^*}$
can be much smaller than in the undriven case.
In the
{\em off-resonance}
situation $(|\epsilon_{n}| > \Delta_{n}$ for all $n$) of
Figs. 3a and 3b, $\sigma_z(t)$  exhibits a complex interference pattern with
\begin{figure}[t]
\epsfig{figure=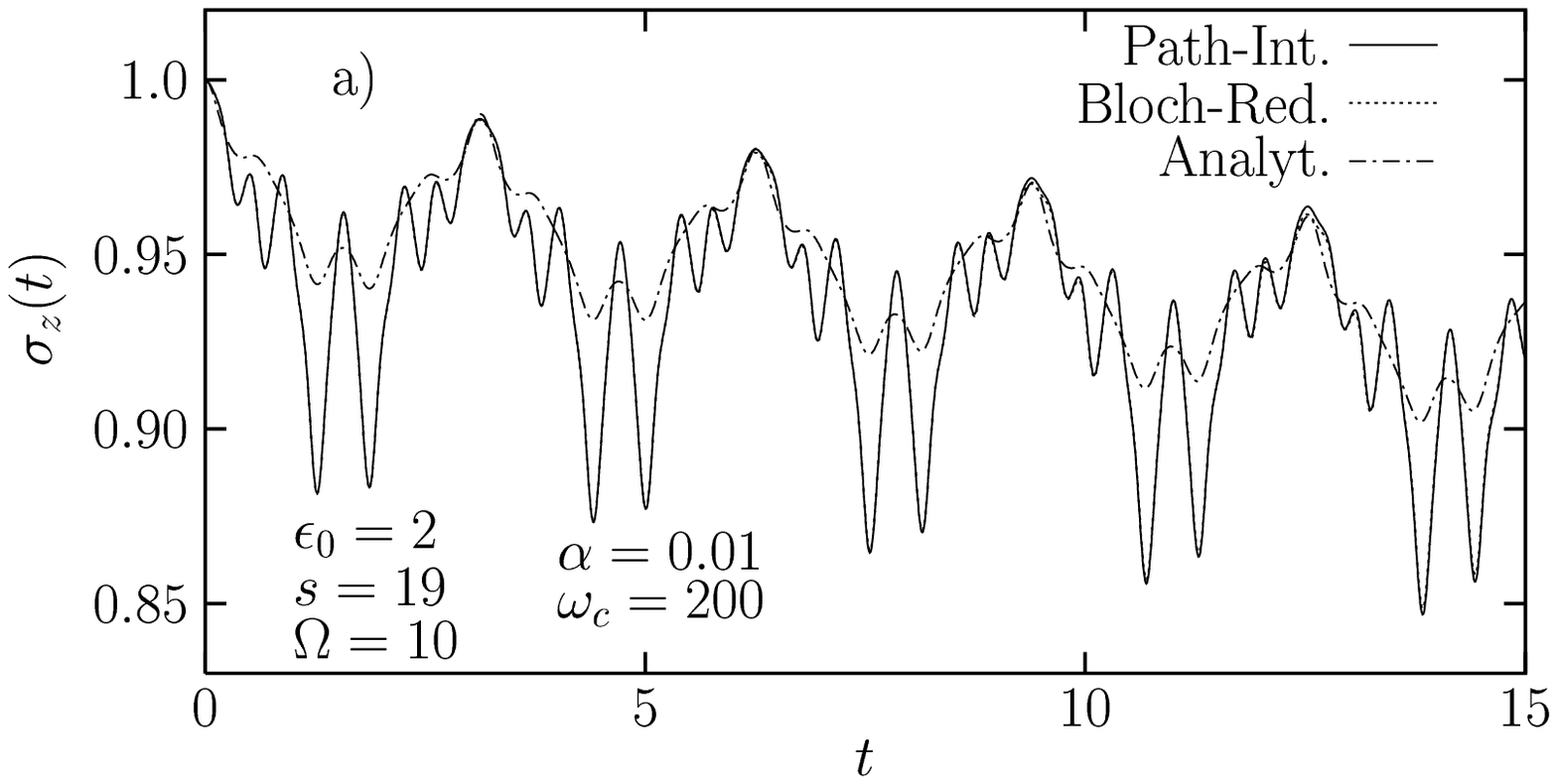,width=8.5cm,height=4cm,angle=0}
\end{figure}
\vspace{-7mm}
\begin{figure}
\epsfig{figure=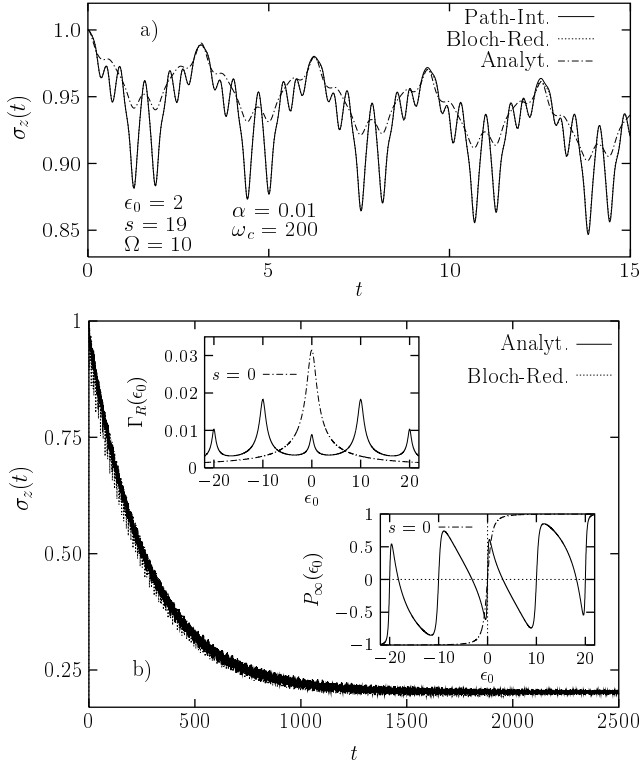,width=8.5cm,height=6cm,angle=0}
\caption{{\em Controlling tunneling.}
In the presence of an {\em off-resonance}
no net separation of time scales occurs
and the population $\sigma_z(t)$ shows a complex interference
pattern (Fig. 3a). - Note that the numerical solutions of Bloch-Redfield
and path-integral equations  coincide within  linewidth.
The TSS dynamics is dominated by an {\em incoherent}
decay towards its asymptotic
limit (Fig. 3b), so that quantum coherence is lost.
The incoherent decay rate $\Gamma_R$, however, can be strongly
diminished. This is demonstrated in the upper left inset where
the {\em photon assisted  decay rate} $\Gamma_R$ is plotted {\it vs.}
the dc-bias $\epsilon_0$. It
exhibits characteristic resonance
peaks at multiple integers of the driving frequency $\Omega$.
These peaks are shifted replicas of the dc-driven ($s =0$)
rate with different weights. Thus,
a suitable chosen bias can enhance or suppress
the decay of populations.
Finally, the lower right inset shows the {\em averaged nonequilibrium
 }  population difference $P_\infty$.
 It exhibits a nonmonotonic  dependence on the dc-bias when combined
with a high frequency field. For appropriate values of the dc-field a
 population inversion ($P_\infty <0$ when $\epsilon_0>0$,
 and {\it vice versa})  can occur.}
\end{figure}
\vspace{-0.2cm}
quantum coherence suppressed.
Moreover, the decay
towards the nonstationary equilibrium value
occurs on a much longer time scale as compared to
the case with $s=0$.
This result, observed recently in \cite{Stockburger}, can be
understood via close inspection of the upper left inset in Fig. 3b, where the
averaged decay rate $\Gamma_R$ is plotted {\it vs.} $\epsilon_0$.
For the chosen parameters
the decay rate is strongly diminished.
Moreover, the lower right inset depicts the averaged nonequilibrium value
$P_\infty$ {\it vs.} the dc-bias $\epsilon_0$. Here, photon assisted
tunneling rules the
possible {\em inversion} of asymptotic population
(i.e. $P_\infty < 0$, for $\epsilon_0 > 0$, and {\it vice versa}).

In concluding we maintain that in the perturbative regime  ($\alpha \ln
(\omega_c/\Delta)\ll 1$) it is numerically advantageous to evaluate the weak
coupling  tunneling dynamics by use of the nonstationary Markovian
Bloch-Redfield equations, as compared to the non-Markovian path  integral GME.
We find numerically  perfect agreement as depicted with Fig. 1 - 3. Within the
time scale of tunneling we find {\it no} observable non-Markovian effects.
On physical
grounds, the same remarks apply to 
the time evolution of the full density matrix.
Note, however, that for the scaling
regime (i.e. cutoff $\omega_c \rightarrow\infty$) Bloch-Redfield theory
increasingly fails; it then is necessary to correct even the path integral GME
with an additional, concocted  renormalization scheme \cite{Sigmax}. 
Finally, our novel analytical
scheme may prove prominent in order  to optimize quantum coherence.

\centerline{* * *}
Financial support was provided  by
the Deutsche Forschungsgemeinschaft (HA1517/14-3, HA1517/19-1;
L.H. and P.H.),
by the German-Israel-Foundation
(G-411-018.05/95; I.G. and P.H.),    the A. v.
Humboldt Foundation (I.G.), and  by  the European Community (M.G.).

\end{multicols}

\vspace{-0.5cm}
\begin{references}
\vspace{-1.5cm}

\bibitem{Leggett87}
A.~J. Leggett {\it et al.,}
Rev. Mod. Phys. {\bf 59}, 1 (1987).

\bibitem{Weiss93}
U. Weiss,
{\em Quantum Dissipative Systems}
(World Scientific, Singapore 1993, Second Edition 1999).

\bibitem{physrep}
M. Grifoni and P. H\"anggi,
{\it Driven Quantum Tunneling},
Phys. Rep. {\bf 304}, 219 (1998).


\bibitem{quantComp}
A. Steane, Rep. Prog. Phys. {\bf 61}, 117 (1998), and Refs. therein;
J. Preskill, Physics Today {\bf 52} (6), 24 (1999).

\bibitem{Gri96}
M. Grifoni, M. Sassetti and U. Weiss,
Phys. Rev. E {\bf 53}, R2033 (1996).

\bibitem{Sigmax}
M. Grifoni, M. Winterstetter and U. Weiss,
Phys. Rev. E {\bf 56}, 334 (1997).

\bibitem{Nakajima}
S. Nakajima,
Prog. Theor. Phys. {\bf 20}, 948 (1958);
R. Zwanzig,
J. Chem. Phys. {\bf 33}, 1338 (1960); for a review see: J. T. Hynes and
J. M. Deutch, Physical Chemistry, An Advanced Treatise, {\bf XIB}, chapt. 11,
 729-836 (Academic Press, New York, 1975).


\bibitem{Goy}
Yu. Dakhnovskii,
Phys. Rev. B {\bf 49}, 4649 (1994);
I.~A. Goychuk, E.~G.Petrov and V. May,
Phys. Rev. E {\bf 52}, 2392 (1995);
Chem. Phys. Lett. {\bf 353}, 428 (1996); M. Morillo and R.~I. Cukier,
 Phys. Rev. B {\bf 54}, 13962 (1996).

\bibitem{rem2}This equivalence holds only for the diagonal
elements of the reduced density matrix considered in the tunneling basis.

\bibitem{Neu}
P. Neu and J. Rau,
Phys. Rev. E {\bf 55}, 2195 (1997).

\bibitem{Argyres}
P.N. Argyres and P.L. Kelley,
Phys. Rev. {\bf 134}, A98 (1964).

\bibitem{Igor95}
(a) I.~A. Goychuk,
Phys. Rev. E {\bf 51}, 6267 (1995);
(b) E.~G. Petrov and V.~I. Teslenko,
Theor. Math. Phys. {\bf 84}, 986 (1990).

\bibitem{Bloch57}
F. Bloch,
Phys. Rev. {\bf 105}, 1206 (1957);
A. G. Redfield,
IBM J. Res. Develop. {\bf 1}, 19 (1957).

\bibitem{Aslangul86}
C. Aslangul, N. Pottier and D. Saint-James,
J. Physique (Paris) {\bf 47}, 757 (1986).

\bibitem{footnote}
Small field-induced corrections to the dephasing rates
$\Gamma_n$ have been disregarded.

\bibitem{Stockburger}
J. T. Stockburger,
Phys. Rev. E {\bf 59}, R4709 (1999).

\end{references}
\end{document}